\documentclass[aps,preprint,superscriptaddress]{revtex4}

\usepackage{times}
\usepackage{amsmath,bm,amsfonts}
\usepackage{dcolumn} % Align table columns on decimal point
\usepackage{graphicx}
\usepackage{latexsym}
\usepackage{color}

\begin{document}

\title{Competition between structural distortion and magnetic moment
  formation in fullerene C$_{20}$}

\author{Myung Joon Han} 
\affiliation{Department of Physics, University of California, Davis, One
  Shields Avenue, Davis, California 95616, USA}

\author{ Gunn Kim}
\affiliation{Department of Physics and BK21 Physics Research
  Division, Sungkyunkwan University, Suwon 440-746, Korea}

\author{Jae Il Lee}
\affiliation{Department of Physics, Inha University, Incheon 402-751, Korea}

\author{Jaejun Yu}
\affiliation{CSCMR and Department of Physics \&\ Astronomy, Seoul National
  University, Seoul 151-747, Korea}
\email[Corresponding author. Electronic address: ]{jyu@snu.ac.kr}

\date{\today }

\begin{abstract} 
  We investigated the effect of on-site Coulomb interactions on the
  structural and magnetic ground state of the fullerene C$_{20}$ based on
  density-functional-theory calculations within the local density
  approximation plus on-site Coulomb corrections (LDA+$U$). The total
  energies of the high symmetry ($I_{h}$) and distorted ($D_{3d}$)
  structures of C$_{20}$ were calculated for different spin
  configurations. The ground state configurations were found to depend on
  the forms of exchange-correlation potentials and the on-site Coulomb
  interaction parameter $U$, reflecting the subtle nature of the
  competition between Jahn-Teller distortion and magnetic instability in
  fullerene C$_{20}$. While the non-magnetic state of the distorted
  $D_{3d}$ structure is robust for small $U$, a magnetic ground state of
  the undistorted $I_{h}$ structure emerges for $U$ larger than 4 eV when
  the LDA exchange-correlation potential is employed.
\end{abstract}

\pacs{71.20.Tx, 75.75.+a, 61.48.-c}

\maketitle

\section{Introduction}

Since the discovery of fullerene C$_{60}$ \cite{C60} various research
activities on alternative fullerene structures of carbon materials have
flourished due to their noble physical properties and potential
applications as next generation electronic devices.  The smallest
fullerene C$_{20}$ has been actively studied after its first successful
production by Prinzbach et al.~\cite{Nature}. Although there have been
many theoretical predictions on its intriguing properties such as
superconductivity, \cite{Miyamoto} vibrational modes,
\cite{Saito-PRL,Saito-PRB} and transport \cite{Roland,Otani,Yamamoto}
along the line of fullerene-based molecular electronics,
\cite{Park,Pasupathy} C$_{20}$ has been a challenge for theorists since
long before its experimental synthesis.  For instance, the stable
structural configuration of C$_{20}$ has not been clearly understood. The
effort to determine the stable structure and relative stability of
C$_{20}$ isomers, e.g., ring, bowl, and cage, has raised a question on the
contribution of electron correlation energy to the structural
stability\cite{Parasuk,Zhang,Grossman,Galli}.  One of the difficulties in
predicting the most stable C$_{20}$ isomer lies in the treatment of the
exchange-correlation (XC) energy. It was shown that
density-functional-theory (DFT) calculations gave qualitatively different
results depending on the XC functional such as local density approximation
(LDA), generalized gradient approximation (GGA), and Hatree-Fock
(HF). Even the quantum Monte Carlo (QMC) results were different from LDA,
GGA, HF and hybrid DFT\cite{Grossman}.

The prediction of the ground state atomic structure of fullerene C$_{20}$
has suffered the similar problem. The high symmetry $I_{h}$ structure of
the C$_{20}$ cage is known to be unstable with respect to the Jahn-Teller (JT)
distortion.  According to Galli et al.~\cite{Galli}, both LDA and GGA
predicted $D_{3d}$ as the symmetry of the most stable Jahn-Teller
distorted cage structure, which is consistent with the tight-binding
molecular-dynamics (TBMD) result by Yamamoto et
al.~\cite{Yamamoto}. However, the hybrid DFT calculation by Saito and
Miyamoto \cite{Saito-PRL} determined C$_{2h}$ to be the most stable one,
while $C_{2}$ or $D_{5d}$ were chosen by the HF calculations\cite{Galli}.
Sometimes even the same kind of calculations made by different groups have
given different predictions (see, for example, Ref.~\cite{Galli}). As
Grossman et al. pointed out, this small molecular carbon system, C$_{20}$,
demonstrates the importance of the highly accurate treatment of electron
correlations\cite{Grossman}.

Recently Lin et al.~\cite{Lin-PRB,Lin-JPCM} took a different perspective
to this issue. By considering the fullerene C$_{20}$ as a superatom with
20 valence electrons, they set up a tight binding model taking account of
the strong hybridization of $p_{\pi}$-orbitals with $p_{\sigma}$, which
arises from the extreme curvature of the C$_{20}$ cage.  Expecting the
important role of the on-site Coulomb interaction, they carried out QMC
and exact diagonalization (ED) calculations for the Hubbard model
hamiltonian for C$_{20}$.  As a result, it was found that C$_{20}$ with
both $I_{h}$ and $D_{3d}$ structures undergoes a magnetic-to-non-magnetic
and non-magnetic-to-magnetic transitions for $I_{h}$ and $D_{3d}$,
respectively, as $U/t$ increases\cite{Lin-PRB,Lin-JPCM}.

To investigate the effect of the on-site Coulomb interaction in connection
with its ground state atomic and magnetic structures, we performed LDA+$U$
and GGA+$U$ calculations \cite{LDA+U} and calculated optimized structure
under the symmetry constraint of $I_{h}$ and $D_{3d}$. Our result showed
that the non-magnetic $D_{3d}$ structure is more stable than magnetic
$I_{h}$ for small values of $U$ less than 4 eV regardless of the choice of
the XC functionals, which is consistent with previous works
\cite{Yamamoto,Galli}. However, when $U>4$ eV, the LDA$+U$ calculations
predict a magnetic state with $\mu=2 \mu_{\mathrm{B}}$ to be a ground
state, which is different from GGA$+U$. This result indicates that there
are subtleties in the competition between the Jahn-Teller structural
instability and the magnetic instability enhanced by Coulomb
correlations. Further the dependence on the exchange-correlation
functionals suggests the importance of electron correlation effects in
determination of the C$_{20}$ ground state.

\section{Computational Methods}

\subsection{LDA+$U$ Method}
In order to describe on-site Coulomb interactions for the carbon $2p$ orbitals
,which are relatively localized in a small-sized
cluster like C$_{20}$, we employed the
LDA$+U$ (GGA$+U$) method \cite{LDA+U,Han-LDA+U}. In our LDA$+U$
(GGA$+U$) approach, the total energy functional is written as
\cite{Han-LDA+U}
\begin{equation}
\label{eq1}
E_{{\rm LDA}+U ({\rm GGA}+U)}=E_{\rm LDA (GGA)} + E^{0}_{U} - E^{U}_{\rm dc},
\end{equation}
where $E_{LDA (GGA)}$ is the LDA (GGA) energy functional and
$E^{0}_{U}$ is given by a spherically averaged screened-Coulomb energy
$U$ and an exchange energy $J$:
\begin{equation}
 \label{eq2}
 E^{0}_{U} = 
        \frac{1}{2}\sum_{\alpha}
        U_{\alpha}
        \sum_{\sigma mm'}
        n^{\sigma}_{\alpha m} n^{-\sigma}_{\alpha m'}
     + \frac{1}{2}\sum_{\alpha}
        (U_{\alpha}-J_{\alpha}) 
        \sum_{\sigma,m\neq m'}
        n^{\sigma}_{\alpha m} n^{\sigma}_{\alpha m'},
\end{equation}
where $\sigma$ is the spin index, $\alpha\equiv (ilp)$ with the site
index $i$, the angular momentum quantum number $l$, and the
multiplicity number of the radial basis function
$p$. $n^{\sigma}_{\alpha m}$ is an eigenvalue of the occupation number
matrix.  Here the $U$- and the $J$-values are assumed to depend only
on the index $\alpha$, but to be independent of the azimuthal quantum
number $m$, which is regarded as a simplification of the Hartree-Fock
theory by using a spherical average. The double-counting term
$E^{U}_{\rm dc}$ can be written as
\begin{equation}
E^{U}_{\rm dc} = \frac{1}{2}\sum_{\alpha} U_{\alpha}
N_{\alpha}(N_{\alpha}-1) - \frac{1}{2}\sum_{\alpha} J_{\alpha}
\sum_{\sigma} N_{\alpha}^{\sigma}(N_{\alpha}^{\sigma}-1),
\end{equation}
where $N_{\alpha}^{\sigma}=\sum_{m}n^{\sigma}_{\alpha m}$ and
$N_{\alpha}=N_{\alpha}^{\uparrow}+N_{\alpha}^{\downarrow}$.
 Therefore, $E_{U} \equiv E^{0}_{U}-E^{U}_{\rm dc}$
reads
\begin{equation}
  E_{U} = 
        \frac{1}{2}\sum_{\alpha}
        (U_{\alpha}-J_{\alpha}) 
        \sum_{\sigma}
        \left\{
        {\rm Tr}({ n_{\alpha}^{\sigma} })
        -  
        {\rm Tr}({ n_{\alpha}^{\sigma} n_{\alpha}^{\sigma}})
        \right\},
\end{equation}
and it is the term that describes the on-site Coulomb interactions.
The effective $U$-value, defined by $U=U_{\alpha}-J_{\alpha}$, can be
treated as a numerical parameter, and it is clear from Eq.~(4) that
LDA$+U$ (GGA$+U$) method is equivalent to LDA (GGA) in $U \rightarrow
0$ limit. We used the previously proposed `dual' representation
\cite{Han-LDA+U}, which is proven to be successful for various
materials \cite{Han-App,Han-App2,Han-App3}.

We considered the on-site Coulomb interaction $U$ for the
carbon $2p$ orbitals, which are relatively localized in a small-sized
cluster like C$_{20}$.  In our LDA$+U$ and GGA$+U$ calculations, we
used the previously proposed `dual' representation \cite{Han-LDA+U},
which is proven to be successful for various materials
\cite{Han-App,Han-App2,Han-App3}.

\subsection{Computation Details}
We performed cluster calculations based on the DFT by employing a
linear-combination-of-localized-pseudo-atomic orbitals (LCPAO) as a
basis set.   Ceperley-Alder \cite{CA,PZ} and
Perdew-Burke-Ernzerh \cite{PBE} type exchange-correlation energy
functional were adopted for LDA and GGA calculation, respectively. We
used double valence orbitals as a basis set which were generated by a
confinement potential scheme with the cutoff radius of 5.0
{a.u.}. Troullier-Martins type pseudo-potentials with a partial core
correction were used to replace the deep core potentials by
norm-conserving soft potentials in a factorized separable form with
multiple projectors.  The real space grid techniques were used with
the energy cutoff of 160 Ry in numerical integrations and the solution
of the Poisson equation using fast Fourier transformations (FFT). In
addition, the projector expansion method was employed to accurately
calculate three-center integrals associated with the deep neutral atom
potential with $L_{\rm max }=6$ and $N_{\rm rad}=4$. All the DFT
calculations were performed using our DFT OpenMX code\cite{OpenMX}.

\section{Result and Discussion}

\begin{figure}[htbp]
  \begin{center}
    \includegraphics[width=10cm]{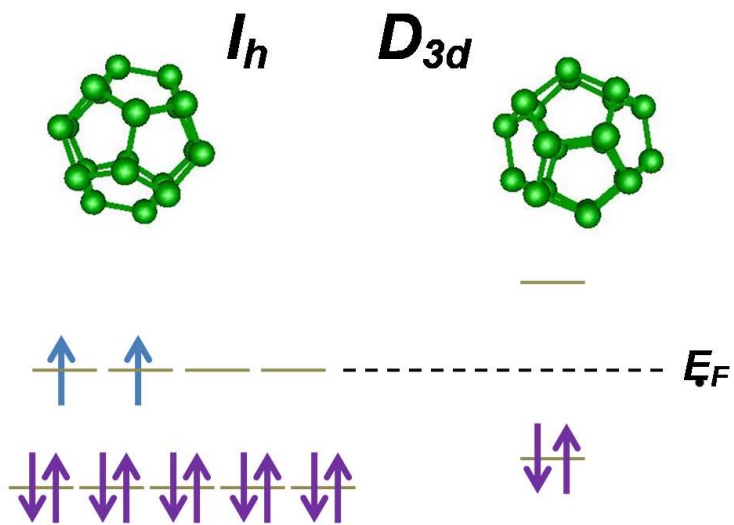}
  \end{center}
   \caption{(Color online) H{\"u}ckel's molecular orbital levels for
     the highly symmetric $I_h$ structure and the Jahn-Teller
     distorted $D_{3d}$ structure of C$_{20}$. Up and down arrows
     indicate up and down spin electrons, respectively. The horizontal
     dashed line represents the Fermi level.
 \label{fig:C20_MO}}
\end{figure}

Figure~\ref{fig:C20_MO} shows a schematic diagram of the H{\"u}ckel's
molecular orbital levels around the Fermi energy for the magnetic $I_h$
and the Jahn-Teller (JT) distorted $D_{3d}$ structure of fullerene C$_{20}$. In
the highly symmetric $I_h$ structure, the
highest-occupied-molecular-orbital (HOMO) states are four-fold degenerate
and partially filled by two valence electrons. Due to the electronic
degeneracy, the $I_h$ structure C$_{20}$ is Jahn-Teller active and can
undergo a structural distortion. Consequently the Jahn-Teller distorted
$D_{3d}$ structure leads to a finite gap in between the HOMO level and the
lowest-unoccupied-molecular-orbital (LUMO) states. Since the HOMO state is
occupied by a singlet pair, the $D_{3d}$ ground state becomes
non-magnetic. On the other hand, the $I_{h}$ C$_{20}$ ground state can be
magnetic due to a kind of the Hund's rule coupling present in this
`superatom', where the magnetic exchange energy between the HOMO electrons
favors the magnetically aligned spin state of the undistorted $I_{h}$
structure.

\begin{figure}[htbp]
  \begin{center}
    \includegraphics[width=10cm]{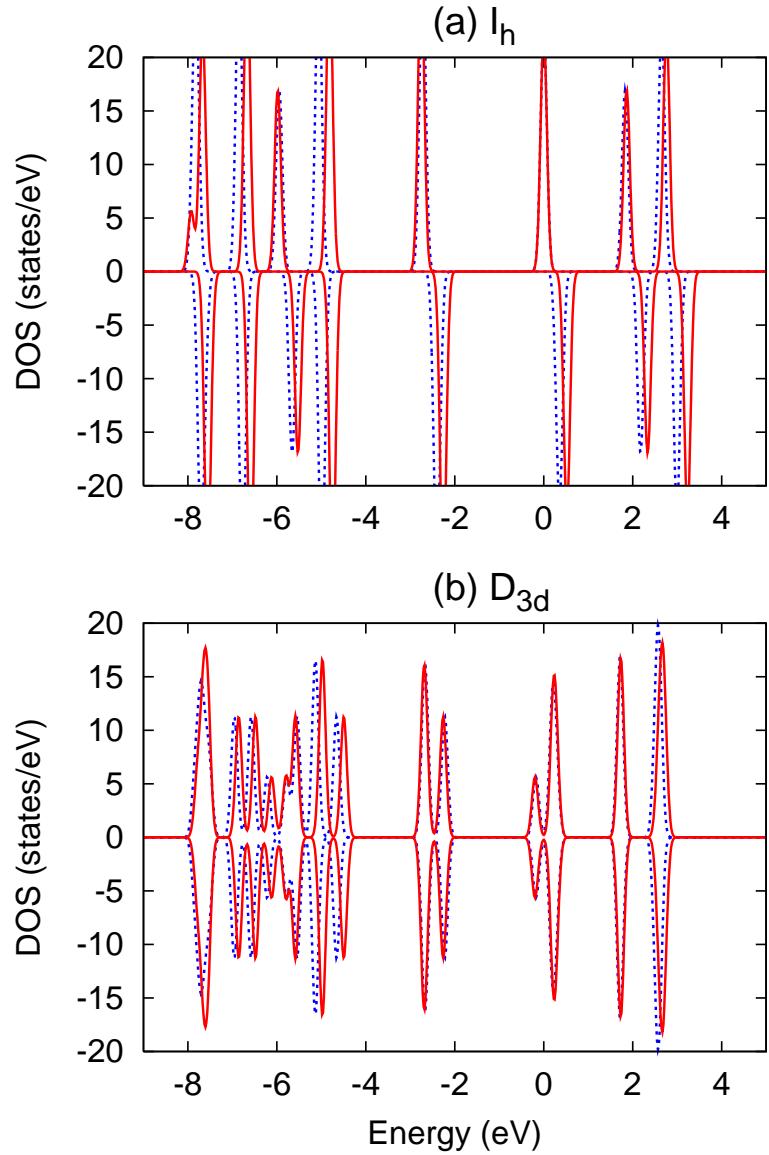}
  \end{center}
   \caption{(Color online) Total density-of-states (DOS) for (a)
    $I_{h}$ and (b) $D_{3d}$ structure C$_{20}$ fullerene. Up and down
    pannels refer to the up-spin and down-spin states, and the dotted
    blue and solid red lines correspond to $U=0$ and 2 eV,
    respectively. The DOS plots were drawn with the Gaussian
    broadening of 0.1 eV, and the Fermi energy is set at zero.
 \label{fig:DOS-both}}
\end{figure}

The calculated density-of-states (DOS) in Fig.~\ref{fig:DOS-both} clearly
shows the different electronic structure of these two structures.  The
dotted blue and solid red lines correspond to the $U=0$ and $U=2$ eV
calculation, respectively.  The Fermi level of $I_h$ C$_{20}$ is located
at the middle of a majority spin peak (Fig.~\ref{fig:DOS-both}(a)),
whereas $D_{3d}$ has a finite HOMO-LUMO gap (Fig.~\ref{fig:DOS-both}(b)),
which is consistent with the H{\"u}ckel's molecular orbital pictures in
Fig.~\ref{fig:C20_MO}.  While the electronic states of the JT-distorted
$D_{3d}$ structure are hardly affected by the change of on-site $U$ values
from 0 to 2 eV, the exchange split of the undistorted $I_{h}$ structure is
significantly enhanced even for the moderate on-site interaction of $U$= 2
eV. Considering the enhanced exchange interactions, we tried to explore a
possible contribution of the on-site Coulomb interactions to the
stabilization of the magnetic $I_{h}$ structure over that of the JT
structural distortion.

\begin{figure}[htbp]
  \begin{center}
    \includegraphics[width=10cm]{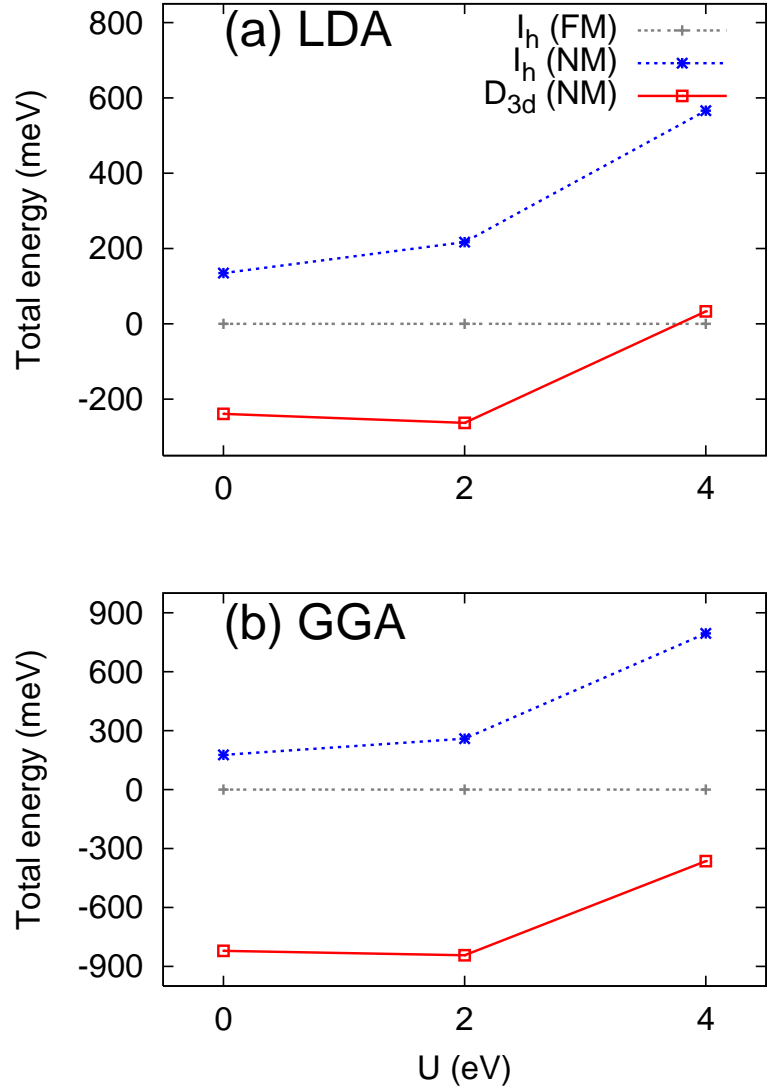}
  \end{center}
   \caption{(Color online) Total energy curves as a function of $U$.
   The calculated energies of NM-$I_{h}$ (dotted blue) and NM-$D_{3d}$
   (solid red) are plotted with respect to that of FM-$I_{h}$ (dotted
   gray) in meV unit.
   \label{fig:EvsU}}
\end{figure}

Figure~\ref{fig:EvsU} shows the calculated total energies by LDA$+U$
(Fig.~\ref{fig:EvsU}(a)) and GGA$+U$ (Fig.~\ref{fig:EvsU}(b)) as a
function of $U$. For comparison, we also calculated the total energies of
the non-magnetic (NM) phase of $I_{h}$. The total energies of FM-$I_h$,
NM-$I_h$, and NM-$D_{3d}$ are represented by the dotted grey, dotted blue,
and solid red lines, respectively.  The total energy of FM-$I_h$ is set at
zero as a reference.  For the both LDA and GGA calculations with $U=0$,
the NM-$D_{3d}$ configuration was found to be the ground state and the
NM-$I_h$ configuration is the highest in energy, which is consistent with
the JT structural instability of $I_{h}$ discussed above. GGA predicts the
larger stability of the non-magnetic NM-$D_{3d}$ over the magnetic
FM-$I_{h}$. The stabilization energy of the NM-$D_{3d}$ over FM-$I_h$ and
NM-$I_h$ is 239 and 374 meV, in LDA, whereas 821 and 998 meV in GGA, ,
respectively. (See the Table~\ref{table} for detailed information.) The
relative order of stability among the different configurations remains the
same even with $U=2$ eV. 

\begin{figure}[htbp]
  \begin{center}
    \includegraphics[width=7cm]{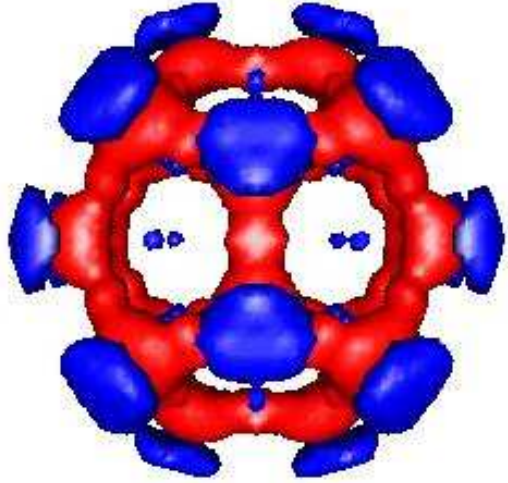}
  \end{center}
   \caption{(Color online) Iso-surface plot of the charge density
    difference of FM $I_h$ between $U=0$ eV and $U=2$ eV.  Blue- and
    red-colored surfaces correspond to electron surplus regions for
    $U=2$ eV and $U=0$ eV, respectively, at the isovalue of $0.7\times
    10^{-3}$ $e\cdot$\AA$^{-3}$.
 \label{fig:CD-diff}}
\end{figure}

It is noted that the $D_{3d}$ structure is more stabilized over $I_h$
in GGA calculation than in LDA for both $U=0$ and 2 eV; the
stabilization energy is more than three times larger than in LDA.
With a larger Coulomb interaction parameter, $U=4$ eV, The total
energy differences between FM-$I_{h}$ and NM-$D_{3d}$ are dramatically
reduced for both LDA$+U$ and GGA$+U$ results. Indeed, in the case of
LDA$+U$, the ground state is changed from the NM-$D_{3d}$ to the
FM-$I_{h}$ configuration as shown in Fig.~\ref{fig:EvsU}. This result
implies that, when $U$ becomes large, the magnetic energy gain with an
enhance exchange interaction can stabilize the symmetric $I_h$
structure against the Jahn-Teller distortion with a lower symmetry
$D_{3d}$ structure.  The strong on-site Coulomb interactions enhance
the electron localization, which leads to the exchange energy gain
thereby contributing to the stabilization of the undistorted
structure. Figure~\ref{fig:CD-diff} shows a charge density difference
between $U=2$ and $U=0$ of FM-$I_h$. The blue and red colors represent
the electron surplus regions of the $U=2$ and $U=0$ eV calculations,
respectively.  Though the magnitude of the charge difference is small,
the difference plot of Fig.~\ref{fig:CD-diff} demonstrates the
tendency of electron localization at each carbon site, which arises
from the electron correlation due to the on-site Coulomb
interactions. The effect of the electron localization can affect the
relatively enhanced exchange coupling among the degenerate HOMO
states, and its magnitude seems to be comparable to that of the JT
distortion. In Fig.~\ref{fig:DOS-both}, it is shown that the JT level
spacing in Fig.~\ref{fig:DOS-both}(b) is comparable to that of the
exchange splitting in Fig.~\ref{fig:DOS-both}(a). Since the JT
distortion lifts up the orbital degeneracy of the HOMO states, the JT
mechanism acts against the formation of magnetic moment in fullerene
C$_{20}$. This result is another example demonstrating the discrepancy
caused by the different XC energy functionals, and further is
consistent with the previous studies \cite{Grossman,Galli} which
emphasized that the structural ground state depend on the type of XC
energy functionals.

It is interesting to compare our LDA$+U$ and GGA$+U$ calculation
results with the recent Hubbard-model-based study by Lin et
al.~\cite{Lin-PRB}.  Starting from an observation that the hopping
parameter $t$ of the C$_{20}$ fullerene should be much smaller than
that of C$_{60}$ but its on-site repulsion $U$ remains the same, which
means that the ratio of $U/t$ is large, they assumed that electron
correlations play an important role in this system and performed ED
and QMC calculation for one-band Hubbard model parameterized by $U/t$,
\begin{equation}
H = -t \sum_{<i,j>, \sigma} (c^{\dagger}_{i\sigma}c_{j\sigma}+H.c)
    +U \sum_{i} n_{i\uparrow} n_{i\downarrow} .
\end{equation}
Their results predicted that the ground state of the $I_{h}$ structure
changes from triplet to singlet at about $U/t\approx 4.10$, while the
ground state of $D_{3d}$ evolves from the singlet (non-magnetic) state
at $U=0$ to a triplet state for $U/t$ larger than 0.5, and then
transit to a singlet state again at $U/t\approx 4.19$. Those
predictions based on the model analysis are in contradiction to our
DFT calculations, where the FM state of the $I_{h}$ structure is
always stable relative to the NM $I_{h}$ state and the JT distorted
$D_{3d}$ structure prefers a non-magnetic ground state. The main
difference between the Hubbard model approach and our DFT calculations
lies on that our DFT calculations take account of the exchange
coupling among the electrons occupying the HOMO states. The explicit
treatment of the degenerate HOMO states is crucial since the energetic
competition between the exchange coupling and the JT structural
distortion is crucial for the degenerate states.

 \section{Conclusion}

We investigated the effect of on-site Coulomb correlations on the structural and magnetic 
properties of the fullerene C$_{20}$ by
carrying out the LDA$+U$ and GGA$+U$ calculations. 
From the comparison of the total energies of the magnetic $I_{h}$
structure and the JT-distorted $D_{3d}$ structure for different values of
the on-site Coulomb interaction $U$, we suggest that the exchange-coupling
driven magnetic instability can possibly override the Jahn-Teller
structural distortion at least for the value of $U=4$ eV within the
LDA$+U$ calculations. Considering the subtle dependence of the ground
state properties of C$_{20}$ on the choice of XC-functional forms as well
as the Hubbard model parameters, more elaborate calculations are required
to resolve the physical picture for the ground state of the fullerene
C$_{20}$, where the magnetic instability due to the Coulomb correlation
effect competes with the structural distortion in the `superatom'
fullerene C$_{20}$.

\begin{acknowledgments}
This work was supported by a Korea Research Foundation (KRF) grant (MOEHRD
KRF-2005-070-C00041). GK acknowledges the support by the post BK
project. The calculations were carried out at the KISTI Supercomputing
Center under the Supercomputing Application Focus Support Program.
\end{acknowledgments}

\newpage

\begin{table}[hd]
  \caption{\label{table} LDA$+U$ and GGA$+U$ total energies (in meV
    unit) of the $I_{h}$ and $D_{3d}$ structures of C$_{20}$ with FM
    and NM spin configurations as a function of $U$. FM-$D_{3d}$
    converges to be NM spin state in $U = 0, 2$ eV (see the text). The
    total energies are given relative to that of FM-$I_{h}$.}
\begin{tabular}{ccrrr} \hline\hline
XC-type & Symmetry  &  $U=0$ eV &  $U=2$ eV & $U=4$ eV\\
\hline
     &$I_{h}$ (FM)  &     0  &     0  &   0 \\
LDA  &$I_{h}$ (NM)  &   135  &   217  & 566 \\
     &$D_{3d}$ (NM) & $-239$ & $-263$ &  33 \\
\hline
     &$I_{h}$  (FM) &     0  &     0  &   0 \\
GGA  &$I_{h}$  (NM) &   177  &   259  & 795 \\
     &$D_{3d}$ (NM) & $-821$ & $-843$ & $-364$\\
\hline\hline
\end{tabular}  
\end{table}

\newpage

\newpage

\newpage

\newpage

\end{document}